\documentclass[twocolumn,aps,showpacs]{revtex4}
\usepackage[latin9]{inputenc}
\usepackage{textcomp}
\usepackage{amsmath}
\usepackage{amssymb}
\usepackage{graphicx}

\makeatletter

\providecommand{\tabularnewline}{\\}

\@ifundefined{textcolor}{}
{%
 \definecolor{BLACK}{gray}{0}
 \definecolor{WHITE}{gray}{1}
 \definecolor{RED}{rgb}{1,0,0}
 \definecolor{GREEN}{rgb}{0,1,0}
 \definecolor{BLUE}{rgb}{0,0,1}
 \definecolor{CYAN}{cmyk}{1,0,0,0}
 \definecolor{MAGENTA}{cmyk}{0,1,0,0}
 \definecolor{YELLOW}{cmyk}{0,0,1,0}
}

\usepackage{color}

\def\NOT(#1,#2){\OneQubitGate(#1,#2){$X$}}

\makeatother

\begin{document}

\title{Single NV Centers as Sensors for Radio-Frequency Fields}

\author{Jingfu Zhang and Dieter Suter\\
 Fakultaet Physik, Technische Universitaet Dortmund,\\
 D-44221 Dortmund, Germany}
\begin{abstract}
We show that a single electron spin can serve as a sensor for radio-frequency
(RF) magnetic fields. The longitudinal and transverse components of
the RF field can be extracted from the phase acquired during free
evolution of the spin coherence. In our experimental demonstration,
a single electron spin of an NV center in diamond serves as an atomic-size
of two components of an RF field.
\end{abstract}

\date{\today}

\pacs{03.65.Ta,07.55.Ge, 76.30.Mi}

\maketitle
{\it Introduction.}--Quantum sensing can be defined as the use of the quantum
properties of a probing system (sensor) for measuring physical quantities,
such as temperature, time, electric and magnetic fields \cite{RevModPhys.89.035002}.
Quantum systems that can be used as sensors include ensembles of nuclear
spins \cite{https://doi.org/10.1002/mrm.21624}, atomic vapors \cite{5778937,Jiang:2020aa},
trapped ions \cite{RevModPhys.87.1419,doi:10.1126/science.abi5226},
Rydberg atoms \cite{Fan_2015}, superconducting circuits (e.g. SQUIDs)
\cite{PhysRev.140.A1628,PhysRevB.83.134501,doi:10.1063/1.2354545}
and nitrogen-vacancy (NV) centers in diamond, either in the form of
ensembles or single spins \cite{Rondin_2014,RevModPhys.92.015004,Suter201750,doi:10.1146/annurev-physchem-040513-103659}.

Over the last years, it became evident that developments in different
quantum technologies can generate useful synergies. As an example,
progress in quantum sensing can be supported by algorithms and concepts
developed for quantum information, where the quantum bits (qubits)
are used for processing information \cite{nielsen,Stolze:2008xy}.
The information content of a quantum state is the essential property
for optimal performance and can be used to determine fundamental limits
to the sensitivity of a specific sensing modality \cite{PhysRevApplied.14.024088}.
In both fields, the information must be protected against unwanted
environmental noise \cite{RevModPhys.88.041001}, while the interactions
with the environment can be tailored such that the sensor extracts
the targeted information but rejects unwanted perturbations \cite{PhysRevLett.107.230501}.
Techniques like dynamical decoupling (DD) \cite{PhysRevLett.106.240501,suter2012,Balasubramanian2009,article10.1088/1367-2630/ab482d}
or quantum error correction \cite{PhysRevLett.116.230502,PhysRevLett.109.100503,RevModPhys.88.041001}
and extended quantum memories \cite{Zaiser2016,0953-4075-44-15-154003,PhysRevA.87.012301,PhysRevB.84.104417,PhysRevLett.106.240501,PhysRevLett.111.020503}
can enhance the sensitivity of quantum sensing.

Here we focus on using single electron spins from NV centers in diamond
\cite{Suter201750} as quantum sensors for oscillating magnetic fields
i.e., AC fields. The main advantages of the NV centers for quantum
sensing include high sensitivity, precision and spatial resolution
down to atomic scale \cite{Abobeih2019}. These beneficial properties
are associated with the strong interaction between the electron spin
and magnetic fields. Powerfull control operations have been developed
for this sensor, using resonant microwave fields and optical excitation.
Readout is accomplished through efficient single-photon counting techniques.
NV centers have been used as quantum sensors in biological systems,
to provide access and insight into the structure and function of individual
biomolecules and observe biological processes at the quantum level
with atomic resolution \cite{doi:10.1002/anie.201506556}.

NV centers can perform as sensors for both DC and AC magnetometry,
with one or multiple centers, e.g., based on Rabi oscillation or spin-locking
\cite{RevModPhys.89.035002,Rondin_2014,PhysRevLett.110.017602,doi:10.1021/acs.nanolett.1c01165,Wang2015NC,PhysRevApplied.10.034044,PhysRevLett.122.100501,Chen_2013,balasubramanian2008nanoscale,arXiv:2206.08533}.
Most previous works on AC magnetometry were based on pulsed DD or
continuous driving techniques such as spin-locking \cite{Rondin_2014,PhysRevLett.110.017602}.
However, with this approach, the frequency or the strength of the AC
field that can be detected is limited by the Rabi frequencies of the
DD pulses, and the continuous microwave (MW) driving or too many DD
pulses might cause undesired effects, such as MW broadening. Moreover,
these techniques are only sensitive to DC fields or to AC fields in
a very narrow frequency range.

In this Letter, we propose and experimentally demonstrate a different
strategy that does not suffer from these limitations: we encode the
longitudinal and the transverse terms of an AC magnetic field in the
phase of a coherent superposition of different spin states, where
the MW fields are used only to generate and detect the coherence and
therefore do not put any limitations to the fields to be measured.
Moreover, in contrast to existing approaches, our scheme can detect
multiple frequency components simultaneously. The experiments were
performed at room temperature, using a diamond sample isotopically
enriched in $^{12}$C to $99.995$\% \cite{1882-0786-6-5-055601,PhysRevLett.110.240501,doi:10.1063/1.4731778}.

{\it Theory.}--The interaction between the electron spin and the RF field
can be described by the Hamiltonian
\begin{equation}
\mathcal{H}_{e}(t)=(\omega_{z}S_{z}+\omega_{x}S_{x})\sin(\omega_{RF}t+\varphi_{0}).\label{eq:Hame}
\end{equation}
Here the amplitudes of the $z-$ and $x$-component of the RF field
are $\omega_{z}=-2\gamma_{e}B_{1,z}^{rf}$, $\omega_{x}=-2\gamma_{e}B_{1,x}^{rf}$
and $\varphi_{0}$ denotes the initial phase of the RF field. $S_{z}$
and $S_{x}$ denote the spin-1 operators for the electron. We use
a coordinate system where the $z$ -axis is oriented along the symmetry
axis of the NV and the RF field lies in the $xz$-plane. The first
term in Eq. (\ref{eq:Hame}) describes the longitudinal component
of the RF field \cite{nature17404}. It commutes with the static system
Hamiltonian and changes the energy levels and transition frequencies
in first order. The second term couples to the transverse spin component
$S_{x}$. It does not generate a first order shift, but the resulting
second-order effect in the RF field amplitude, which is known as Bloch-Siegert
shift (BSS) \cite{PhysRev.57.522,PhysRevA.98.052354}, also contributes
to a shift of the energy levels. Since the second-order effect also
commutes with the static Hamiltonian, it can be treated independently
of the effect of the longitudinal term.

We start with the longitudinal term, which shifts the energy levels
of the electron spin by $\mathcal{\delta E}_{m}=-\omega_{z}m\sin(\omega_{rf}t+\varphi_{0})$,
where $m$ is the corresponding eigenvalue of $S_{z}$. A coherence
between states $|m\rangle$, $|m'\rangle$ then acquires a phase 
\begin{eqnarray}
\varphi_{z}(\tau_{p}) & = & -\omega_{z}(m-m')\int_{0}^{\tau_{p}}\sin(\omega_{rf}t+\varphi_{0})dt\nonumber \\
 & = & \alpha[\cos(\omega_{rf}\tau_{p}+\varphi_{0})-\cos(\varphi_{0})],\label{eq:phase}
\end{eqnarray}
where $\alpha=(m-m')\omega_{z}/\omega_{rf}$ and $\tau_{p}$ is the
duration of the RF pulse.

\begin{figure}
\centering{}\includegraphics[width=0.8\columnwidth]{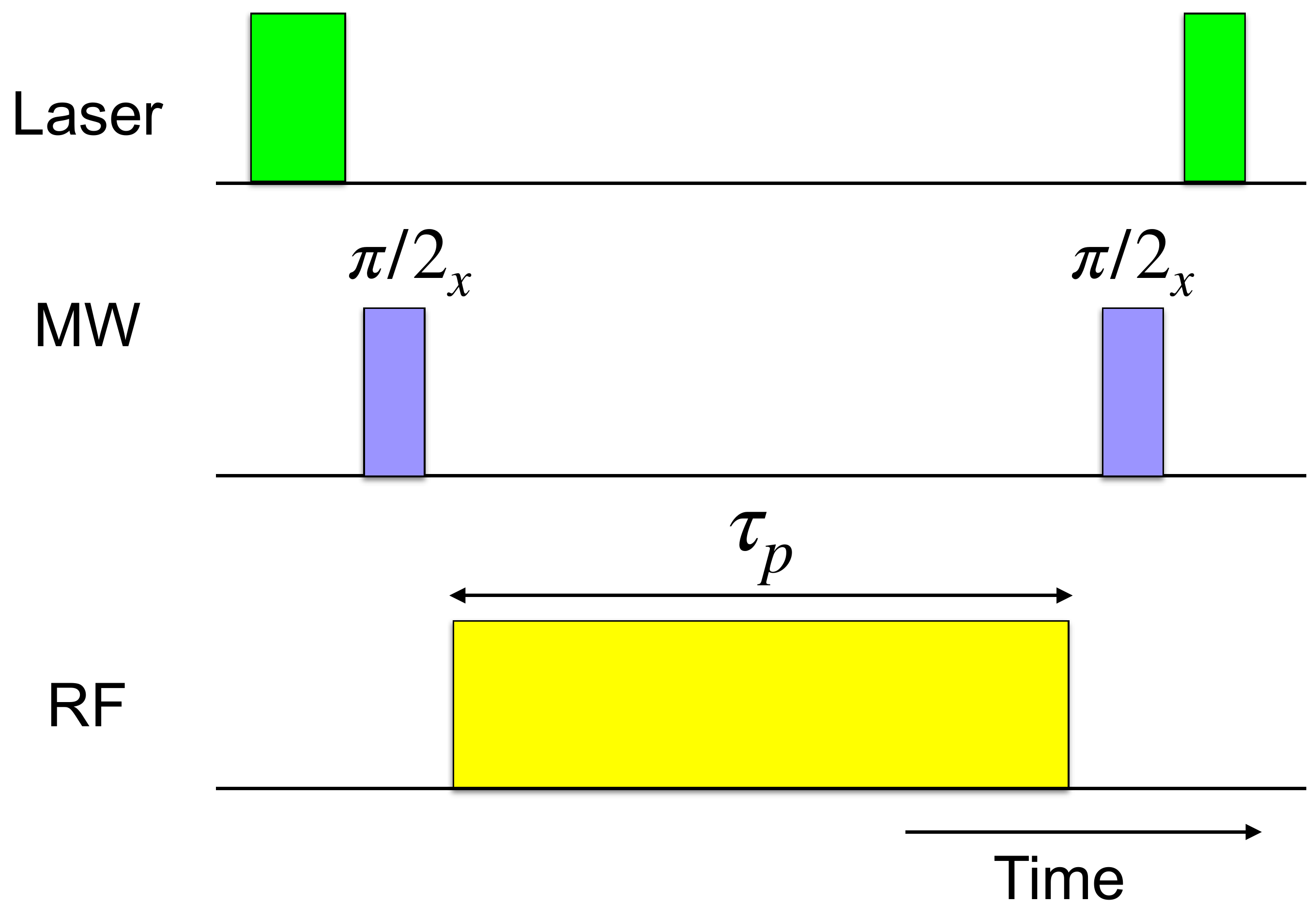}
\caption{Pulse sequence for sensing RF fields using the electron spin. The
electron spin is initialized in state $|0\rangle$ by the first laser
pulse. The first $\pi/2$ MW pulse generates the superposition of
the spin states, and the second $\pi/2$ pulse converts one component
of the coherence to population, which then can be read out by the
second laser pulse. \label{PulseSeq}}
\end{figure}

The effects of the acquired phase can be observed using the pulse
sequence shown in Fig. \ref{PulseSeq}. The initial state of the electron
spin is $|m\rangle$. The first $\pi/2$ pulse generates the superposition
of states $|m\rangle$ and $|m'\rangle$. The second $\pi/2$ pulse
converts part of the coherence to population, which can be read out.
The resulting signal depends on $\varphi_{z}$ as
\begin{equation}
P_{|0\rangle}(\tau_{p})=[1-\cos\varphi_{z}(\tau_{p})]/2.\label{eq:p0-2}
\end{equation}

For spectral analysis, we write $\cos\varphi_{z}(\tau_{p})$ as a
Fourier series by using
\begin{equation}
\cos\varphi_{z}(\tau_{p})=\cos(\alpha\cos\varphi_{0})S_{c}+\sin(\alpha\cos\varphi_{0})S_{s}.\label{eq:Fs}
\end{equation}
Here $S_{c}\equiv\cos(\alpha\cos x)$, $S_{s}\equiv\sin(\alpha\cos x)$
are Fourier series:
\begin{eqnarray}
S_{c} & = & J_{0}(\alpha)+2\sum_{n=1}^{\infty}(-1)^{n}J_{2n}(\alpha)\cos(2nx)\nonumber \\
S_{s} & = & 2\sum_{n=0}^{\infty}(-1)^{n}J_{2n+1}(\alpha)\cos[(2n+1)x]\label{eq:sc-1}
\end{eqnarray}
where $x=\omega_{rf}\tau_{p}+\varphi_{0}$ and $J_{\nu}(z)$ is the
Bessel function of the first kind \cite{Saiko2019}.

For the discussion of the second order effect (BSS), we consider a
transition between two electron states with transition frequency $\Omega_{0}$.
The BSS shifts this resonance frequency by
\begin{equation}
\omega_{BS}=\omega_{x}^{2}/(2\Omega_{0}).\label{eq:BSth}
\end{equation}
The acquired phase of the coherence between states $|m\rangle$ and
$|m'\rangle$ is 
\begin{equation}
\varphi_{x}(\tau_{p})=\omega_{BS}\tau_{p}.
\end{equation}
The effects of the BSS can also be observed using the pulse sequence
in Fig. \ref{PulseSeq}. Eq. (\ref{eq:p0-2}) becomes then
\begin{equation}
P_{|0\rangle}(\tau_{p})=\{1-\cos[\varphi_{z}(\tau_{p})+\varphi_{x}(\tau_{p})]\}/2.
\end{equation}

{\it Experimental demonstration for short signals.}--We choose the electron
states $|m_{S}=0\rangle$ and $|m_{S}=-1\rangle$ for a quantitative
study of these effects, starting with the first-order effect. We first
initialize the electron into $|m_{S}=0\rangle$ with the details given
in section I of the supplementary material (SM). According to Eqs. (\ref{eq:Fs}-\ref{eq:sc-1}),
the first-order effect contributes components at frequencies $n\omega_{rf}$.
We therefore record signals over a few periods $2\pi/\omega_{rf}$.
On this timescale, dephasing effects are small (the dephasing time
of the electron spin $T_{2}^{*}\approx22$ $\mu$s, see section I
of the SM). We use the pulse sequence shown in Fig. \ref{PulseSeq},
with the RF frequency $\omega_{rf}/(2\pi)=2$ MHz \cite{noteZcouple}.

\begin{figure}
\centering{}\includegraphics[width=1\columnwidth]{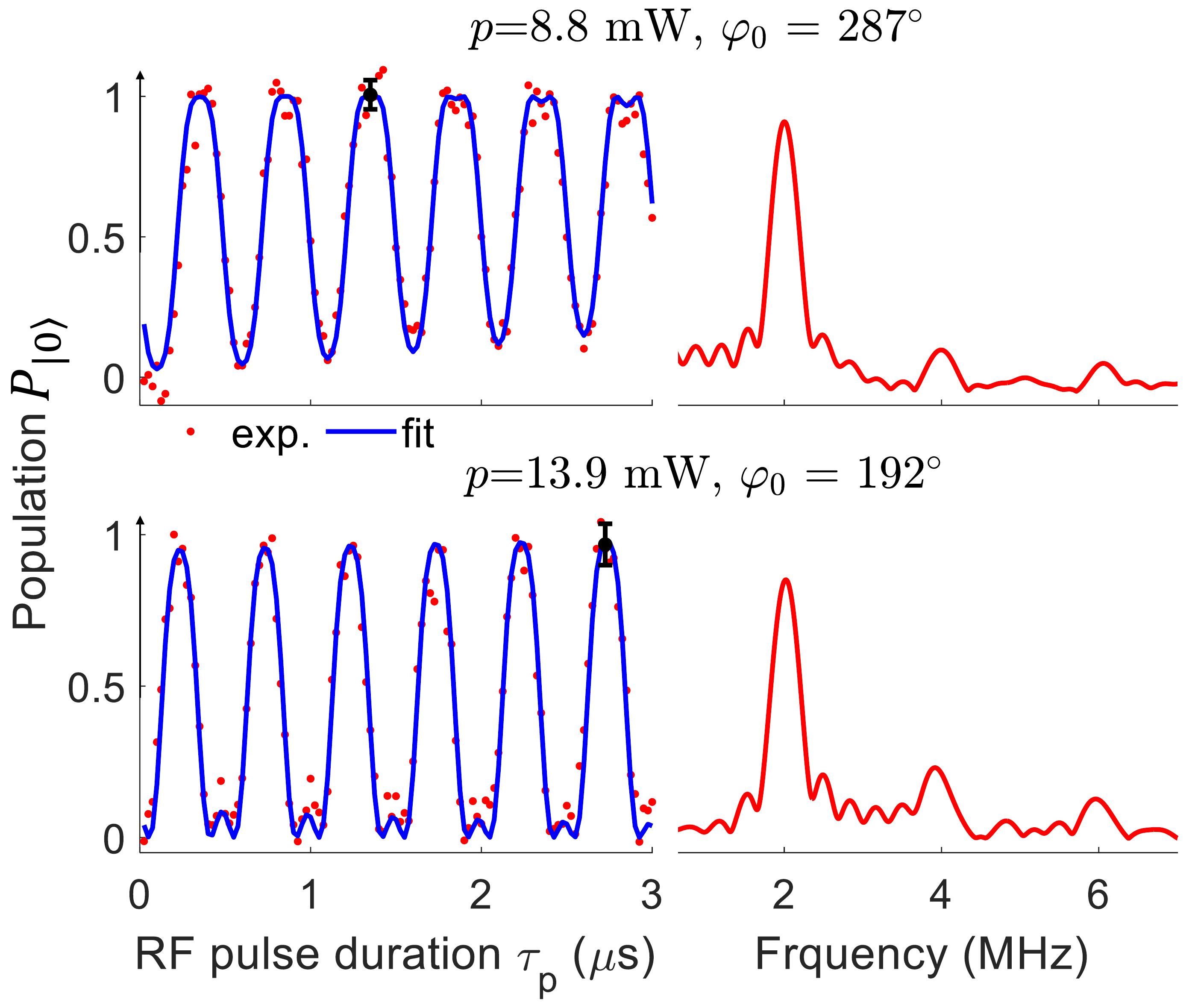}
\caption{Experimental results demonstrating the effect of the longitudinal
component of the RF field on the electron spin at two different RF
powers $p$ of the RF pulses. The left column shows the population
$P_{|0\rangle}$ as a function of the RF pulse duration $\tau_{p}$.
The error bars indicate the photon counting statistics. The dots represent
the experimental data and the curves the fit to function \eqref{eq:p0-2},
where the phase $\varphi_{z}(\tau_{p})$ is given by Eq. \eqref{eq:phi_fit}.
The right column shows the corresponding absolute value spectra. \label{figure_za}}
\end{figure}

Fig. \ref{figure_za} shows the experimental results for two different
RF powers and initial phases $\varphi_{0}$. The experimental time-domain
signals (Fig. \ref{figure_za} left) show the expected periodicity
in $2\pi/\omega_{rf}=0.5$ $\text{\ensuremath{\mu}s}$. After several
periods, the signal deviates from exact periodicity. These deviations
are mostly due to the DC-component of the field and to the second-order
effect, as discussed below. To include them in the fitting function,
we modify $\varphi_{z}(\tau_{p})$ in Eq. (\ref{eq:phase}) to
\begin{equation}
\varphi_{f}(\tau_{p})=\varphi_{z}(\tau_{p})+(\omega_{DC}+\omega_{BS})\tau_{p}+\delta.\label{eq:phi_fit}
\end{equation}

Here $\omega_{DC}$ is the projection of the DC component of the applied
field to the $z$ -axis and $\omega_{BS}$ the second-order contribution
of the AC component, which generates a time-averaged frequency shift
(see Eq. \eqref{eq:BSth} and Ref. \cite{PhysRev.57.522,PhysRevA.98.052354}).
$\omega_{DC}$ and $\omega_{BS}$ have different dependencies on the
RF power level: $\omega_{DC}$ increases with the square root of the
power and $\omega_{BS}$ increases quadratically with $\omega_{x}$
and thus linearly with the power. We can therefore separate the contributions
by evaluating them at different power levels; details are given in
the SM (section IIIE). The constant term $\delta$ appears to be due
to transients generated by switching the RF pulse on and off with
a finite rise time. The values for $\alpha$, $\beta$ and $\delta$
obtained by fitting the experimental data are listed in the SM (section
IIIA). The field amplitudes $\omega_{z}$ at two different power levels
are listed in Table \ref{params}. The ratio between the two measured
values of $\omega_{z}$ is 1.22, consistent with the ratio of the
field strengths $\sqrt{13.9/8.8}=1.26$.

\begin{table}
\begin{tabular}{|c|c|c|}
\hline 
 & $p=8.8$ & $13.9$ (mW)\tabularnewline
\hline 
$\omega_{z}/2\pi$ (MHz) & $2.66\pm0.02$ & $3.24\pm0.12$\tabularnewline
\hline 
$\omega_{x}/2\pi$ (MHz) & $35.2\pm0.7$ & $44\pm1$\tabularnewline
\hline 
$\omega_{DC}/2\pi$ (MHz) & $-0.20\pm0.01$ & $-0.25\pm0.01$\tabularnewline
\hline 
\end{tabular}

\caption{Measured field amplitudes. $\omega_{z}/2\pi$ was obtained from the
experimental data in Fig. \ref{figure_za}, and $\omega_{x}/2\pi$
and $\omega_{DC}/2\pi$ from Fig. \ref{figure_BS}.}
\label{params} 
\end{table}

The periodicity of these data suggests an analysis in the frequency
domain. As shown in the Fourier transforms of the time domain data
in Fig. \ref{figure_za}, peaks appear at integer multiples of the
RF frequency, $n\omega_{rf}$, in agreement with Eq. \eqref{eq:Fs}.
Using the spectra in Fig. \ref{figure_za}, we obtain the values of
$\omega_{z}/2\pi$ consistent with the results obtained from the time-domain
signals. The results are presented in the SM(section IIIB), together
with additional details.

\begin{figure}
\centering{}\includegraphics[width=1\columnwidth]{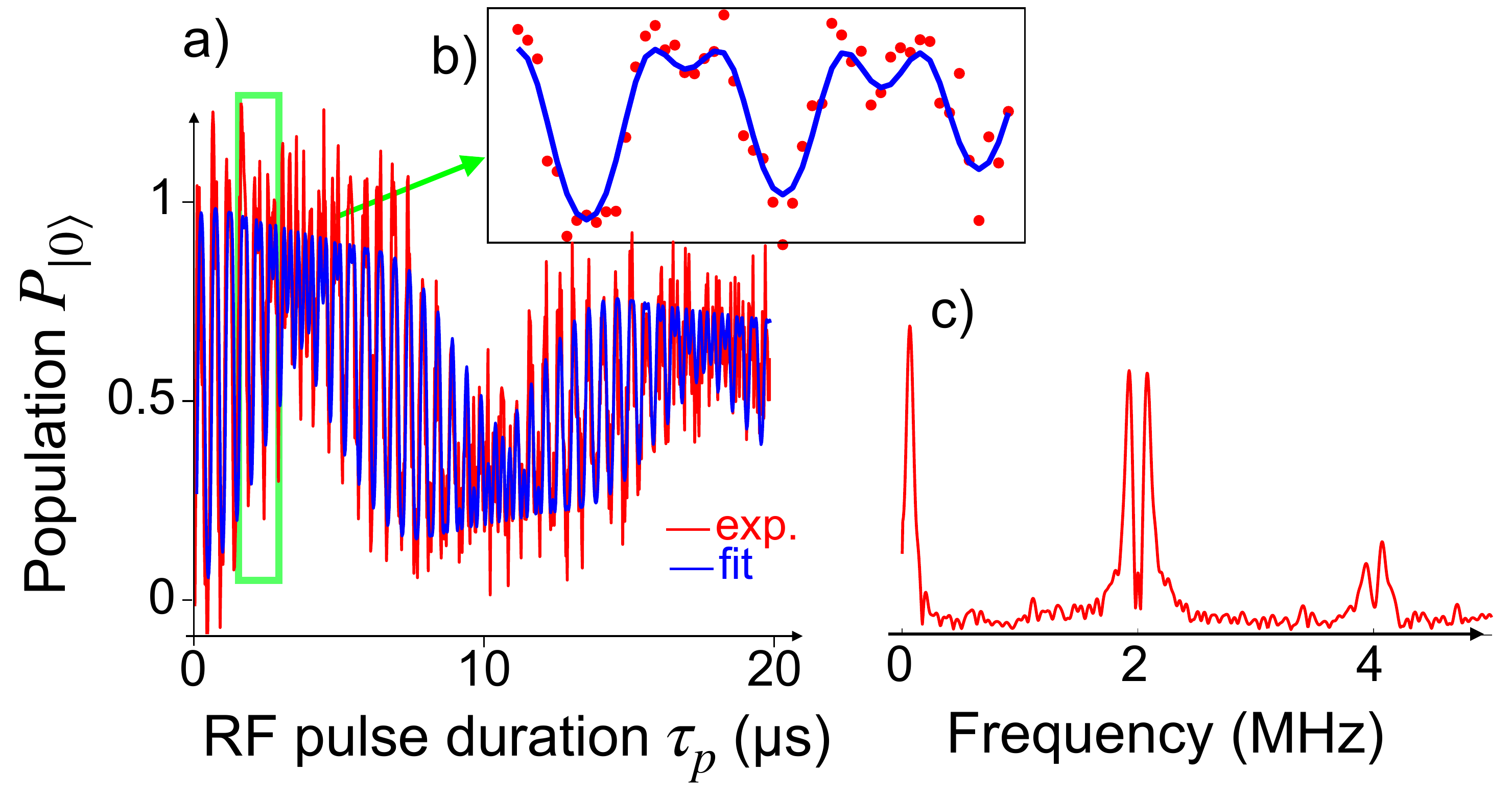}
\caption{Experimental results demonstrating first and second order contributions
to the phase acquired by the electron spins by the pulse sequence
shown in Fig. \ref{PulseSeq}. (a) The population $P_{|0\rangle}$
as a function of the RF pulse duration $\tau_{p}$. The red curve
indicate the experiment data and the blue curves the fit by the function
in Eq. (\ref{eq:fit1}). (b) Details for a short section, where the
experimental data are indicated by filled circles. (c) Absolute value
spectrum obtained from the experimental data in (a). \label{figure_BS-1}}
\end{figure}

{\it Measurement of the second order shift.}--Since second order shifts
are significantly smaller, measuring them requires higher precision
and thus longer signals. Fig. \ref{figure_BS-1} (a-b), shows some
experimental results obtained with the pulse sequence in Fig. \ref{PulseSeq}
and 8.8 mW RF power. The time domain signal can be fitted by 
\begin{equation}
P_{|0\rangle}(\tau_{p})=\{1-e^{-\tau_{p}/T_{2}^{*}}\cos[\varphi_{f}(\tau_{p})]\}/2,\label{eq:fit1}
\end{equation}
where $T_{2}^{*}=22$ $\mu$s, estimated from the FID measurement,
see SM(section I). 

\begin{figure}
\centering{}\includegraphics[width=1\columnwidth]{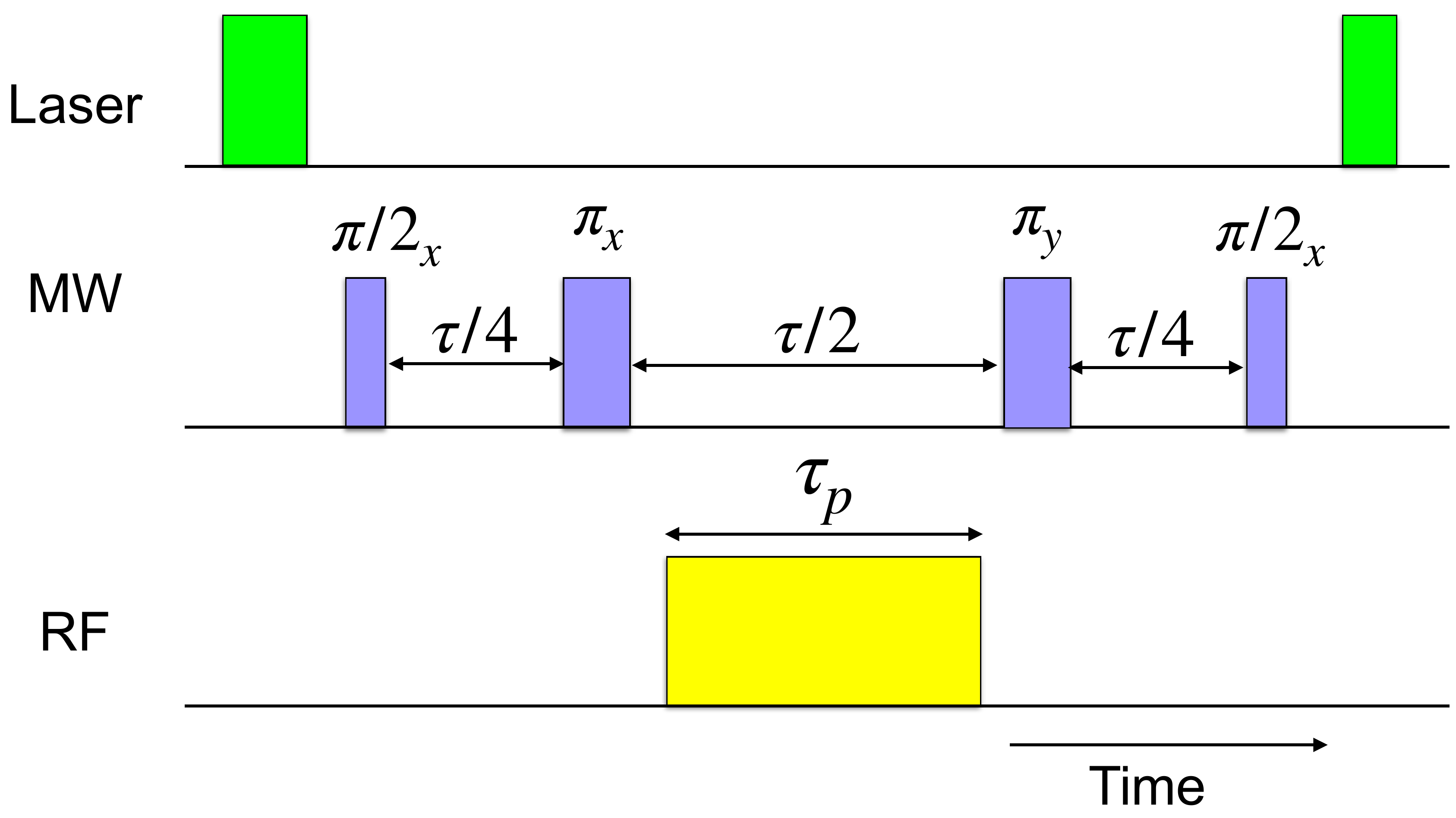}
\caption{Pulse sequence for sensing RF fields over longer time scales. The
two $\pi$ pulses are used to extend the coherence time of the electron
spin. The other pulses are identical to those in Fig. \ref{PulseSeq}.
\label{figure_BSpul}}
\end{figure}

To reduce noise-induced dephasing during these longer measurement
periods, we use DD pulses \cite{suter2012,RevModPhys.88.041001}.
The pulse sequence shown in Fig. \ref{figure_BSpul} includes two
refocusing pulses with a $\pi/2$ phase shift for compensating cumulative
pulse errors \cite{GULLION1990479}. The phase generated by the RF
pulse is not cancelled by the DD pulses, since the RF is applied only
between the two refocusing pulses. The resulting phase is given by
Eq. (\ref{eq:phi_fit}) and it is again transferred into measurable
population by the final $\pi/2$ pulse, resulting in the signal
\begin{equation}
P_{|0\rangle}(\tau_{p})=\{1+\cos[\varphi_{f}(\tau_{p})]\}/2.\label{eq:fit2}
\end{equation}

\begin{figure}
\centering{}\includegraphics[width=1\columnwidth]{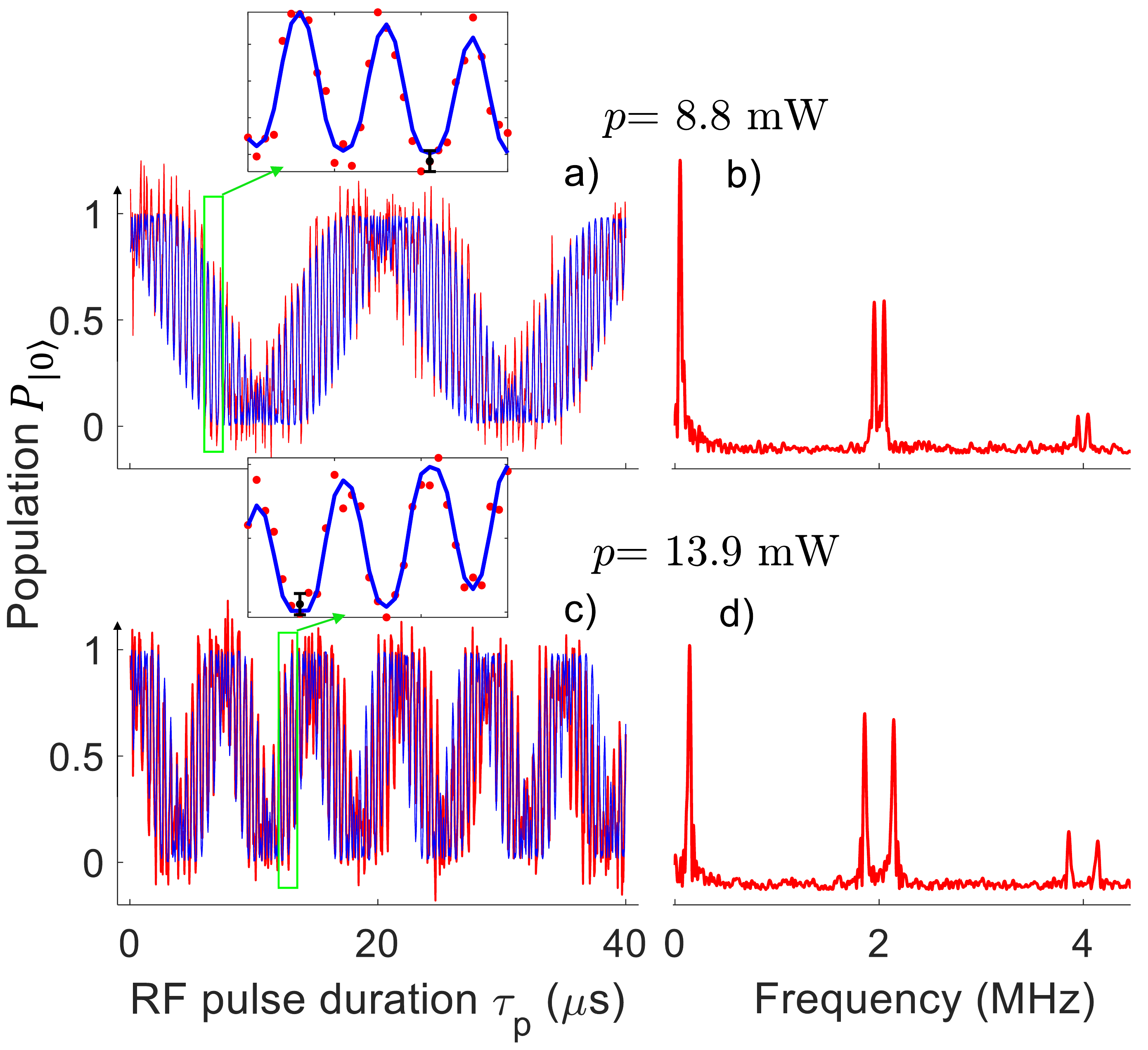}
\caption{Experimental results demonstrating first and second order contributions
to the phase acquired by the electron spins at two different RF powers,
using the pulse sequence shown in Fig. \ref{figure_BSpul}. The left
column shows the population $P_{|0\rangle}$ as a function of the
RF pulse duration $\tau_{p}$. The red curves indicate the experimental
data and the blue curves show the fit to function (\ref{eq:fit2}).
The right column shows the absolute value spectra obtained from the
experimental data on the left. \label{figure_BS}}
\end{figure}
 Fig. \ref{figure_BS}(lhs) shows the measured signal $P_{|0\rangle}(\tau_{p})$,
for two different powers. Compared with the signal in Fig. \ref{figure_BS-1}
(a), the decay due to the dephasing effect is negligible, since the
DD pulses extend the dephasing time up to $1.2$ ms \cite{PhysRevA.98.052354}.
The fast oscillation (period \textless{} 1 \textmu s) is due to the
first order effect covered by the previous section, while the slower
oscillation, whose period decreases drastically when the power level
increases, is the topic of this section. The Fourier transforms of
the time-domain data shown in Fig. \ref{figure_BS} (rhs) contain
peaks at the frequencies $|n\omega_{rf}\pm(\omega_{DC}+\omega_{BS})|/2\pi$,
with $n=0$, 1 and 2. The values of the measured frequencies are listed
in the SM(section IIIC). The calculated field amplitudes $\omega_{DC}$
and $\omega_{x}$ are listed in Table \ref{params}, where Eq. (\ref{eq:BSth})
is used and $\Omega_{0}$ is measured as 2.475151 GHz. These values
agree with those obtained with the simpler pulse sequence in Fig.
\ref{PulseSeq} (for details see SM(section IIIE)), but provide higher
precision and accuracy. From the measured components $\omega_{z}$
and $\omega_{x}$, we can estimate the angle $\theta$ between the
NV axis and the RF field as $\theta\approx86^{\circ}$.

{\it Conclusion.}--This work introduces a protocol for measuring time-dependent
magnetic fields with a large frequency range, starting at zero. It
can detect components parallel as well as perpendicular to the quantization
axis with different sensitivity: the parallel (secular) component
scales linearly with the amplitude of the field while the perpendicular
component contributes in second order. Compared to methods based on
spin-locking \cite{PhysRevLett.110.017602}, our method is valid for
weak and strong RF fields and covers a much larger frequency range.
The experimental implementation is performed using the electron spin
of an NV center in diamond. We consider a system where the field is
magnetic, but it should be equally applicable to electric fields.

{\it Acknowledgments.}--This project has received funding from the European
Union's Horizon 2020 research and innovation programme under grant
agreement No 828946. The publication reflects the opinion of the authors;
the agency and the commission may not be held responsible for the
information.

\bibliographystyle{apsrev}

\begin{thebibliography}{10}
\bibitem{noteZcouple} The specific frequency was chosen to be sufficiently far
from the resonance frequency of the NMR transitions, in
order to avoid effects that are not related to the sensing
via the electron spin.
\end{thebibliography}

\end{document}